# Unambiguous definition of handedness for locally-chiral light

**Ofer Neufeld[1,2] and Oren Cohen[2]**

[1]Max Planck Institute for the Structure and Dynamics of Matter, Hamburg, Germany, 22761.
[2]Physics Department and Solid State Institute, Technion - Israel Institute of Technology, Haifa 32000, Israel.
*Corresponding author e-mail: oneufeld@schmidtsciencefellows.org.

**Abstract**

Synthetic chiral light fields were recently introduced as a novel source of chirality [Ayuso *et al*., *Nat. Phot.* **13**, 866 (2019)]. This locally-chiral light spans a three-dimensional polarization that plots a chiral trajectory in space-time, leading to huge nonlinear chiral signals upon interactions with chiral media. The degree of chirality of this new form of light was defined, characterized, and shown to be proportional to the chiral signal conversion efficiency. However, the sign of the light's chirality – its 'handedness' – has not yet been defined. Standard definitions of helicity are inapplicable for locally-chiral light due to its complex three-dimensional structure. Here, we define an unambiguous handedness for locally-chiral fields and employ it in practical calculations.

## 1. Introduction

Electromagnetic (EM) fields can carry intrinsic structure ranging from orbital angular momentum [1,2], to topological invariants [3–8], and more [9–20]. Recently, a form of twisted light was proposed that carries a new property – 'local-chirality' – a chiral charge embedded in the three-dimensional time-dependent polarization structure of the vector field [21], (which breaks all dynamical improper-rotational symmetries [22]). The degree of chirality (DOC) of this light source was characterized [23], and it was predicted to be highly useful for chirality sensing though nonlinear light-matter interactions in high harmonic generation (HHG) [21,24–26] and photoionization [27]. However, the handedness (also often referred to as helicity) of locally-chiral light has not yet been characterized or well-defined. Due to its complex three-dimensional structure, standard measures for helicity (e.g. ellipticity-helicity [28], time-dependent ellipticity-helicity [29], optical chirality [30–36], spectral-helicity [37], etc.) are inapplicable, or irrelevant. An unambiguous definition of locally-chiral light's handedness is perquisite for further development of applications, as well as for a cohesive and unified formalism.

Here we define an unambiguous handedness for locally-chiral light denoted as 'hDOC'. This definition can be consistently applied to any vector field. The hDOC is based on a unique triple product for a given field, and is odd under parity. It embodies similar principles to those used by the definition of handedness in molecular chemistry employed by the International Union of Pure and Applied Chemistry (IUPAC) [38,39]. It establishes a formal pseudoscalar for characterizing light's local chirality that relies solely on the field's time-dependent polarization structure, without need for further analysis of the interaction of the field with chiral matter.

## 2. hDOC

We first briefly review the DOC of locally-chiral light (henceforth simply denoted DOC for brevity), which is given by [21]:

$$\chi \equiv \frac{\min_{\theta,\phi,\psi,\Delta t}\left\{\int dt \left|\hat{R}_z(\psi) \cdot \hat{R}_x(\phi) \cdot \hat{R}_z(\theta) \cdot \vec{E}(t+\Delta t) + \vec{E}(t)\right|\right\}}{\int dt \left|\vec{E}(t)\right|} \quad (1)$$

here $\chi$ is the DOC of the electric field $\vec{E}(t)$ at a single point in space. The angles $\psi$, $\phi$, and $\theta$, are three Euler angles, $\hat{R}_j$ denotes a rotation operator about the $j$'th axis, and $\Delta t$ is a temporal shift. The operator min{} is a minimization procedure over the three Euler angles and $\Delta t$. The magnetic field is ignored in eq. (1) because its interaction with matter is neglected in the electric dipole-approximation. We also note that for freely propagating laser beams the magnetic and electric fields contribute equally to the local-chirality density. Eq. (1) originates from a geometrical definition of chirality, in accordance with its definition in molecular chemistry and IUPAC - an object is chiral if and only if it cannot be superposed onto its mirror image. Thus, the integral in the numerator of eq. (1) evaluates the subtraction between the electric field vector, $\vec{E}(t)$, and its inverted twin, $-\vec{E}(t)$, while allowing one of the twins the freedom to rotate freely in space-time (by the Euler angles and time shift). In case that one of the twins can be fully superposable onto the other, there exists some set of angles and temporal delay (those minimized over) such that the integral in eq. (1) vanishes (i.e. the minimization procedure yields $\chi=0$). For any other case, one obtains a quantitative measure for how similar the field is to its mirror twin, which defines the DOC as an integrated overlap measure [40].

Notably, $\chi$ in eq. (1) is positive definite, and is normalized between 0 and 200% (due to the denominator). Mathematically, this is a result of the absolute value inside the integral in the numerator. Physically, it is a result of the averaging procedure for the overlap between the field and its mirror twin over time, which can only be done with positive definite values. Thus, $\chi$ in eq. (1) is currently not a pseudoscalar, but a scalar, and cannot describe chiral observables – it is not odd under parity. The missing 'sign' in eq. (1) is the hDOC that will be defined below, which will in turn normalize the DOC from -200% to 200%, just like chiral dichroism.

We first briefly review the definition of handedness in molecular stereochemistry (given by IUPAC), which provides an unambiguous protocol to determine the handedness of chiral molecular compounds. As an illustrative example, we follow the case of bromochlorofluoromethane (CBrClFH), the smallest stable chiral molecule that has one chirality center (here a carbon atom that is bonded to four different atomic substituents). Figure 1(a) presents the protocol for determining the molecule's handedness for this particular example. According to the definition, the substituents around the chiral center are numbered from heaviest to lightest, where the lighter substituents take the lowest priority (in this case hydrogen is the lightest, and is thus denoted by '4'). The molecule is then oriented such that the lowest priority substituent points towards the back of the viewer (into the plane in Fig. 1(a)). An arrow is then drawn from the substituents '1'→'2'→'3'. If the arrow rotates clockwise, the handedness is termed 'R'. Otherwise, it is termed 'S'. Importantly, while at first glance this procedure seems to take an arbitrary geometrical approach for chirality handedness, it actually relies on a rigorous mathematical principle – a pseudoscalar formed by a triple-product. With this procedure, and by aligning the lightest constituent (here '4') away from the viewer, the arrow drawn from '1'→'2'→'3' in fact evaluates the normalized triple product $(\vec{1} \times \vec{2}) \cdot \vec{4}$, which gives $\pm 1$ depending on the handedness of the molecule, and is odd under parity (here $\vec{i}$ is a vector originating from the chiral center to substituent '$i$'). We also note that this definition is unambiguous (though not unique), and is solely a property of the molecule, unlike alternative definitions that rely on the interaction of the molecule with other entities such as light or other molecules.

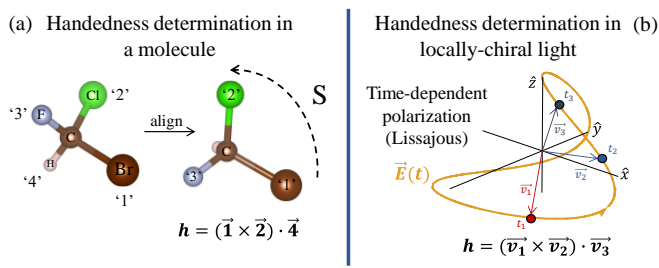

**Figure 1.** (a) Illustration for handedness determination protocol in chiral molecules for the example of CBrClFH. The different substituents around the chiral center are numbered according to their weight, with the lightest group aligned pointing away from the viewer. An arrow is then drawn from the heaviest element towards the second heaviest element, which determines the handedness of the molecule, i.e. whether its 'R' or 'S' (see text). (b) Illustration of main protocol for determination of the hDOC in locally-chiral light. An exemplary Lissajous of a locally-chiral field is plotted, which traces a chiral trajectory with its time-dependent polarization. Three unique vectors are chosen according to three unique moments in time, determined by the moment when the field's instantaneous intensity is maximized ($t_1$). The normalized triple-product of these vectors defines the handedness ($h$) of the field.

We now use an analogous approach for defining handedness for locally-chiral light. This is motivated by the geometrical origin of the local-chirality, which is already similar to the origin of chirality in molecules (except that it is extended to the spatio-temporal domain of vector fields). The problem is then essentially reduced to defining a unique triple-product for a given field, $\vec{E}(t)$, and associating its sign as the hDOC. We emphasize that the vectors that comprise the triple product must be absolutely unique for a given $\vec{E}(t)$ (or the definition becomes inconsistent when comparing different light fields). Following the approach in molecules, one could define three unique vectors directly from $\vec{E}(t)$ itself by choosing three unique moments in time: $t_1, t_2, t_3$, with the vectors $\vec{v_1} = \vec{E}(t_1)$, $\vec{v_2} = \vec{E}(t_2)$, $\vec{v_3} = \vec{E}(t_3)$. How to determine those moments in time is a somewhat arbitrary procedure, with many possible conventions (just as in the case of molecules). Nonetheless, once a convention is chosen the definition would be consistent, unambiguous, and unique for a given field.

We shall first define the general rule and then address the pathological cases. Continuing the analogy with molecules, we determine $t_1$, $t_2$, and $t_3$ based on the field's instantaneous intensity, $I(t) = |\vec{E}(t)|^2$, which plays an analogous role to the substituents weights in a molecule. With this in mind, one sets the first moment in time, $t_1$, as the moment that maximizes $I(t)$:

$$I(t_1) \equiv \max\{I(t)\} \tag{2}$$

Next, $t_2$ could be simply defined as an arbitrary, but unique, temporal translation away from $t_1$, e.g. $t_2=t_1+T/3$, where $T$ is the temporal period of the time-periodic part of $\vec{E}(t)$ (which would be the Floquet period in standard cases of coherent laser fields). $t_3$ could similarly be defined as $t_3=t_2+T/4$. From these we obtain the hDOC as:

$$h_{DOC} = \frac{(\vec{v_1} \times \vec{v_2}) \cdot \vec{v_3}}{|(\vec{v_1} \times \vec{v_2}) \cdot \vec{v_3}|} = \frac{(\vec{E}(t_1) \times \vec{E}(t_2)) \cdot \vec{E}(t_3)}{|(\vec{E}(t_1) \times \vec{E}(t_2)) \cdot \vec{E}(t_3)|} \tag{3}$$

, which defines the main branch of the protocol for determining hDOC. An illustration for this protocol can be seen in Fig. 1(b) for an exemplary case of locally-chiral light.

Of course, just as in the IUPAC definition there are various procedures in the flow chart that address pathological cases where the above protocol might fail (e.g. cases where two substituents have the same wight, etc.). In what follows, we too describe all possibilities where the procedures outlined above fail, and present alternative procedures that form a complete protocol.

First, it is possible that there is no single unique moment in time, $t_1$, that maximizes $I(t)$. For instance, there could be several finite moments in time with an equal maximal instantaneous power, such that $I(t_j) = \max\{I(t)\}$ for several values $\{t_j\}$. If the set $\{t_j\}$ is finite, then $t_1$ can be chosen from it, in which case we chose to utilize the behavior of the derivative of $I(t)$, $\partial_t I(t)$. That is, we choose the moment in time that maximizes $I(t)$, and also has the largest $\partial_t I(t)$ out of the set $\{t_j\}$. If the derivatives are also identical for several points, one move on to the next order of derivative, and so on and so forth. If after all considerations, there still remain several points in time that have identical behavior, then that must mean that the field exhibits an internal symmetry structure due to a dynamical symmetry [22] that renders those moments in time identical. In such a case, any moment in time out of those equivalent points can be chosen as $t_1$ as they will all lead to the same hDOC.

On the other hand, if the set $\{t_j\}$ is infinite, i.e. there are an infinite number of moments in time that lead to the same maxima in $I(t)$, a different approach must be taken. In this case, we follow the protocol described above, except that $t_1$ is given by the moment in time that maximizes the field's angular velocity, $|\partial_t \vec{E}(t)|$, that is:

$$|\partial_t \vec{E}(t_1)| \equiv \max\{|\partial_t \vec{E}(t)|\} \tag{4}$$

This approach relies on the rate at which the vector field rotates instead of its power. Notably, this pathological case can never occur in locally-chiral light that is comprised of freely propagating waves, as those cannot have a constant power over time. However, one can imagine, as a thought experiment, a vector field that lives on a sphere (e.g. an arrow of length $r$, $|\vec{E}(t)|=r$, which only changes its direction in time), which falls into the category above – it does not have any unique moment in time that maximizes $I(t)$, but it can still exhibit local-chirality because the trajectory of its time-dependent polarization can be chiral. If this approach fails, one moves on to the next order of derivative, and so on and so forth.

Second, we outline possible issues in determining $t_2$. Given that $t_1$ is uniquely defined, and $t_2$ is defined as a temporal translation from $t_1$, there could only be one possible issue: due to a coincidence (or due to a physical degeneracy), the obtained $\vec{E}(t_1)$ is parallel to $\vec{E}(t_2)$, or we have that $\vec{E}(t_2) = 0$. In such cases, the two moments in time do not define unique independent vectors. To solve this, one infinitesimally translates time by positive $\delta t$, such that $t_2=t_1+T/3+\delta t$, until a $\delta t$ is found for which the vectors are no longer parallel or until $\vec{E}(t_2) \neq 0$. Such a moment in time is guaranteed to exist if the vector field is not linearly-polarized. If it is linearly polarized, then the DOC in eq. (1) is zero, and hDOC is irrelevant.

Lastly, we outline possible issues in determining $t_3$. Similarly to $t_2$, the only case for which the above procedure fails is if by a coincidence we obtain that $\vec{E}(t_3)$ is parallel to either $\vec{E}(t_1)$ or to $\vec{E}(t_2)$, or that $\vec{E}(t_3) = 0$. If that is the case, the vectors are not independent. To address this, one infinitesimally translates time by positive $\delta t$, such that $t_3=t_2+T/4+\delta t$, until a $\delta t$ is found for which the vectors are no longer parallel. If the DOC in eq. (1) is nonzero, such a moment in time is guaranteed to exist (because otherwise the field polarization is contained in a plane and it is achiral, and hDOC is irrelevant).

Overall, this protocol establishes an unambiguous definition for the handedness of locally-chiral light, and is summarized in the flow chart in Fig. 2.

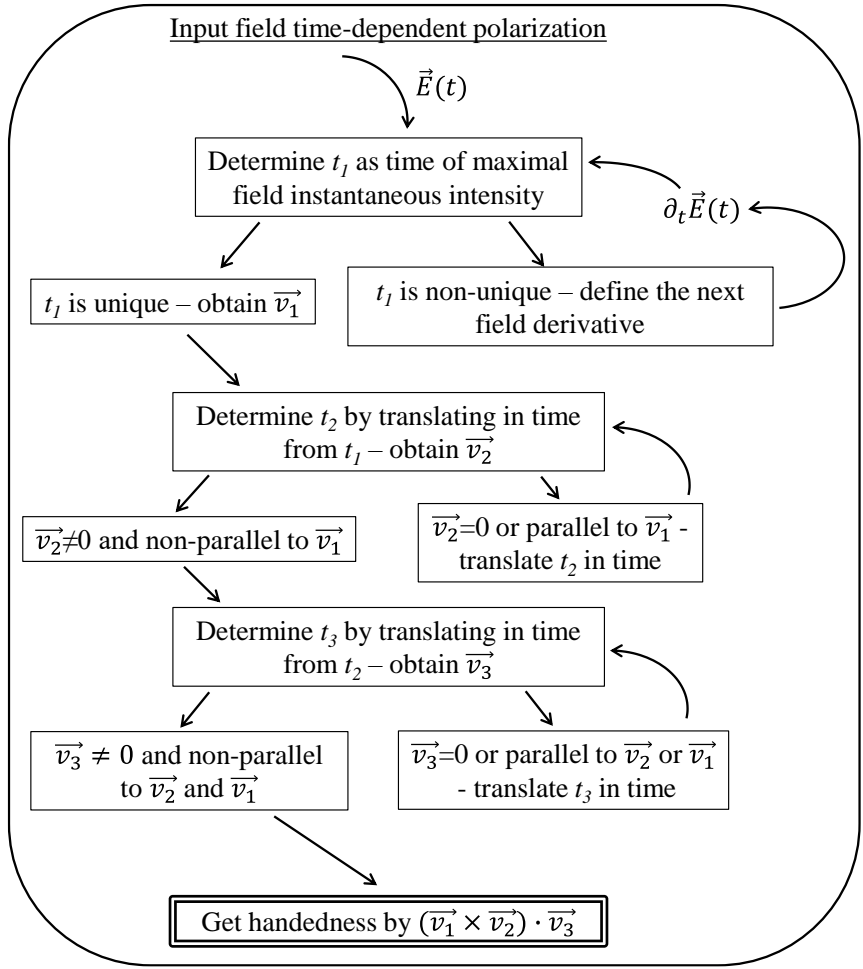

**Figure 2.** Flow chart describing the complete protocol for determining locally-chiral light's hDOC.

We now highlight some noteworthy points: (i) the protocol is unique for a given field, $\vec{E}(t)$. (ii) $h_{DOC}$ is a pseudoscalar that is odd under parity. (iii) This protocol only depends on generic properties of $\vec{E}(t)$ such as the fundamental period, $T$, the maximal power, and the field's space-time characteristics. Because of this, simple scaling of field power or frequency do not change the handedness, as desired. (iv) This protocol is arbitrary in the sense that many other conventions could have been employed (as is often the case with helicity), but it can be consistently applied to any vector field. (v) With it, one can extend the DOC to include light's helicity:

$$\chi \equiv h_{DOC} \frac{\min_{\theta,\phi,\psi,\Delta t}\{\int dt \,|\hat{R}_z(\psi) \cdot \hat{R}_x(\phi) \cdot \hat{R}_z(\theta) \cdot \vec{E}(t+\Delta t) + \vec{E}(t)|\}}{\int dt\,|\vec{E}(t)|} \qquad (5)$$

## 3. Implementing hDOC for analyzing global chirality

In this section we utilize the new definition for exploring locally chiral light fields that are spatially-dependent. When the electric field vector is spatially dependent, i.e. $\vec{E} = \vec{E}(t,\vec{r})$, a unique DOC and helicity can be defined in every point in space, $\vec{r}$. If the light field's phase is rapidly oscillating in space, this can affect the light-matter interaction of such a field with chiral media, as was recently studied in HHG [21,24–26]. Thus, determining light's DOC (including handedness) over a finite region in space is crucial for understanding the importance of so-called global-chirality [21] and 'polarization-chirality' [24], and its impact on chirality spectroscopy. Figure 3 explores several cases that have recently been studied in the literature, where the field's DOC changes across a one-dimensional interaction region. We now analyze several different cases:

(A) We explore a case where locally-chiral light is constructed from two non-collinear ω-2ω bi-chromatic beams with a fixed relative phase of $\pi$, where the ω-components of the beam create elliptical polarization in a plane, while the 2ω-components are transversely linearly-polarized. This type of light was shown to be 'globally-chiral' (i.e. having a constant helicity across the interaction region), yielding large enantio-sensitive signals in the far field [21]. Figure 3(a) shows that our definition of hDOC correctly reconstructs the uniform handedness of the light across the interaction region, just as was determined in ref. [21] using chiral correction functions in the interaction picture (note that there is some noise in the DOC calculation in regions where the field power is very low). We emphasize that no clear definition for handedness of a general field was put forward in ref. [21], and the global-chirality was determined there ad-hoc based on interaction-related properties. Our approach on the other hand can be straightforwardly applied to study global-chirality in various setting and with different fields in a consistent manner, and with relying solely on the time-dependent structure of the laser fields.

(B) We explore another case where tailored light is constructed from ω-2ω elliptically-polarized beams that are non-collinearly aligned, but where each beam is elliptically polarized and the beam configuration is asymmetric in space. This case was shown to be highly locally-chiral [23], leading to strong chiral signals in photoionization that allow all-optical enantio-separation [27] and allowing multi-chiral separation [26]. On the other hand, it was recently argued that the enantio-sensitive HHG response generated by it in the far-field should vanish due to an underlying microscopic-macroscopic dynamical symmetry in the vector field [41]. We show here that this field indeed has a rapidly oscillating chiral handedness (see Fig. 3(b)). The field's handedness changes sign 8 times in every spatial period across the interaction region, while the absolute value of the degree of chirality remains roughly constant. Thus, this beam configuration could be useful for chiral spectroscopy only in a highly confined interaction region relative to the beam parameters. Notably, the field changes its handedness even in spatial positions where its DOC is nonzero. This is a natural property expected from handedness, which further highlights the non-uniqueness of the definition. The physical consequence of this effect is that the exact position in which a field changes its handedness is not necessarily physically meaningful, and can depend on the particular definition.

(C) We study a case where the optical set-up is similar to case (A), yet with different relative phases leading to so-called 'polarization-chirality' [24]. The light in this case is not 'globally-chiral' as in (A), yet still yields a large enantio-sensitive signal in the far-field [24] owing to its internal spatial structure that is asymmetric (where one handedness is more pronounced than the other or structured along a preferred direction). We validate these results using the new measure for chirality – hDOC (see Fig. 3(c)) – the DOC of the light is asymmetric with a more favorable

'R' handedness along the *x*-axis. Using the new measure, one can consistently compare different fields to uncover which should lead to a stronger chiral response in the far field. For instance, we show in Fig. 4(d) that similar 'polarization-chirality' structure can be obtained even when removing the 2ω components from one of the beams, simplifying possible experimental realizations.

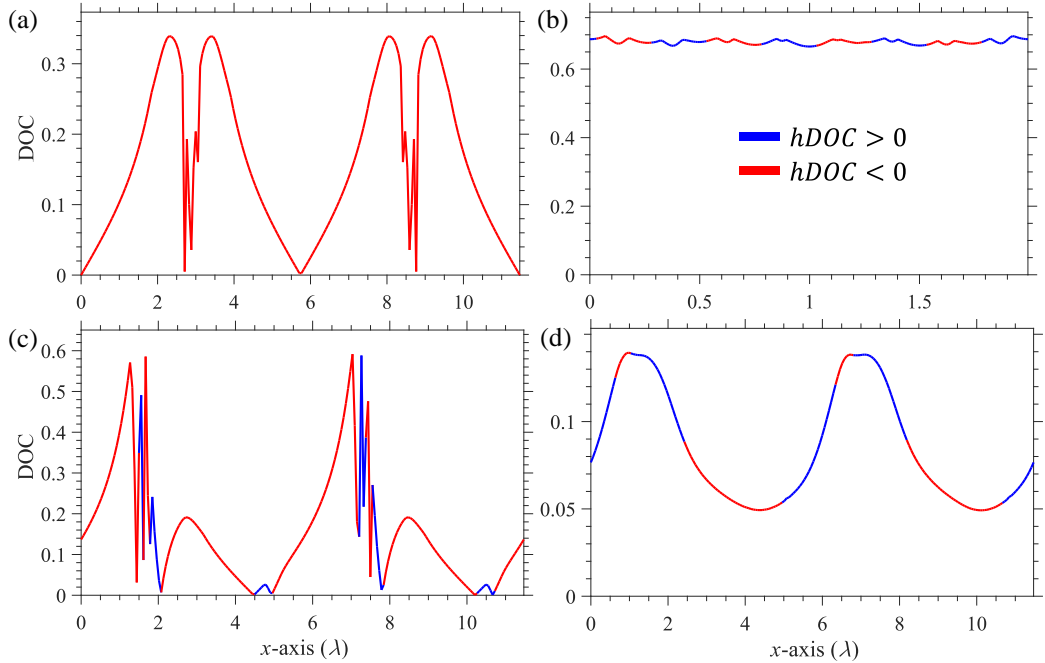

**Figure 3.** Numerical calculations of DOC and hDOC for exemplary cases of locally-chiral light. In all sub-plots the hDOC of light is plotted across a one-dimensional interaction region where two non-collinear light beams overlap (and their relative phases oscillate). hDOC oscillates with the relative phases leading to various structures in the DOC that correspond to enhanced or weakened chiral responses. In all figures, blue refers to light with 'R' handedness, while red to light with 'S' handedness. (a) hDOC corresponding to case (A) described in the main text, where field parameters are taken from Figure 5 in ref. [21]. hDOC maintains a constant handedness across the interaction region. (b) hDOC corresponding to case (B) described in the main text, where field parameters are taken as in Figure 2 in ref. [27]. (c) hDOC corresponding to case (C) described in the main text, where field parameters are taken from Figure 2 in ref. [24]. (d) hDOC for a non-colinear field geometry similar to that in (C), but where the 2ω beam intensity in one of the arms is set to zero (the other beam parameters are taken randomly – the ellipticities are taken as 0.9:0.5:0.2 for the ω:ω:2ω beams respectively, the beam amplitudes ratios are 1:0.5:1 for the ω:ω:2ω beams respectively – but the result of chirality asymmetry along the *x*-axis is generally insensitive to these parameters).

## 4. Summary

To summarize, we have defined an unambiguous handedness for locally-chiral light. Our definition relies on similar principles to those used in chiral molecules, and is applicable to any vector field. We have used this definition for formulating a pseudoscalar that describes the chirality of locally-chiral light, which relies only on intrinsic properties of the light field (and does not require knowledge of light-matter interactions). Lastly, we employed it for exploring some cases of recent interest in chiral light-matter interactions, showing that our definition agrees with previous ad-hoc descriptions and is useful for quickly analyzing experimental set-ups. Most importantly, our work establishes a clear protocol by which one can analyze chiral light-matter interactions and plan future experiments.

## Acknowledgements

We thank O. Smirnova and A. F. Ordonez for insightful discussions. This work was supported by the by the Israel Science Foundation (grant no. 1781/18). O.N. gratefully acknowledge the support of the Adams Fellowship Program of the Israel Academy of Sciences and Humanities, support by the Alexander von Humboldt foundation, and from a Schmidt Science Fellowship.